# Hybrid fiber links for accurate optical frequency comparisons

Won-Kyu Lee,[1,3,4] Fabio Stefani,[1,2] Anthony Bercy,[2,1] Olivier Lopez,[2] Anne Amy-Klein,[2] and Paul-Eric Pottie[1,*]

[1]*Laboratoire National de Métrologie et d'Essais–Système de Références Temps-Espace (LNE-SYRTE), Observatoire de Paris, CNRS, UPMC, Paris, France*

[2]*Laboratoire de Physique des Lasers, Université Paris 13, CNRS, Villetaneuse, France*

[3]*Korea Research Institute of Standards and Science, Daejeon 34113, South Korea*

[4]*University of Science and Technology, Yuseong, Daejeon 34113, South Korea*

*\* paul-eric.pottie@obspm.fr*

## *Abstract*

We present the first experimental demonstration over a 43-km-long urban fiber network of a local two-way optical frequency comparison, which does not require any synchronization of the measurements. It was combined with a regular active-noise compensation on another parallel fiber leading to a very reliable and robust frequency transfer. This hybrid scheme enables us to investigate the major limiting factors of the local two-way comparison. We analyze the contribution to the phase noise of the recovered signal by the interferometers at local and remote places. By using the ability of this set up to be injected by a single laser or two independent lasers, we measure the contribution to the long-term instability by the demodulated laser instabilities. We show that a fractional frequency instability level of $1 \times 10^{-20}$ at 10 000 s can be obtained with this simple set-up after 43-km-long propagation in an urban area.

## 1. Introduction

The technique of phase-compensated optical fiber links has been developed rapidly over the last decade enabling transfer and comparison of optical frequency over continental scales exceeding 1000 kilometers [1–4]. Clock frequency comparison by optical fiber is not only useful for time-frequency metrology, but it also provides a powerful experimental tool for many applications, such as test of variation of fundamental constants, relativistic geodesy, and dark matter search [2, 5].

The optical frequency transfer by the method of active noise cancellation (ANC) [6] has wide applications, such as remote high-resolution and accurate spectroscopy [7–9], very long baseline interferometry [10], and remote narrow-linewidth light source [8, 11, 12]. However, if only comparison of optical frequency is needed, the recently-proposed two-way method [13–15] is a good alternative with the advantage of simplified experimental apparatus without active components. In this method, the fiber noise is eliminated from the frequency comparison signal by post-processing the data obtained synchronously at both sites in the same way as for two-way satellite time and frequency transfer [16].

The two-way optical phase comparison was first introduced in [13], where a Sagnac interferometer configuration was used to compare ultra-stable lasers through 47-km-long fiber link. This was further investigated in [14, 15] with a novel scheme of local two-way (LTW) comparison, where the fiber noise can be suppressed in real-time, by using measurements from one link end only, either at the local side or the remote side. Firstly introduced for a uni-directional configuration using a pair of parallel fiber over an urban network [14], it was presented then for a fully bi-directional configuration over 25-km fiber spool [15]. The LTW method has advantages compared to the conventional two-way (CTW) method. First, since only local measurement data are used, we avoid the need for the synchronization of the measurements between the local site and the remote site to remove the fiber noise; and a real-time frequency comparison is possible. Second, the results of the two LTW can be compared and cross-checked in post-processing, and the CTW observable can be formed. It allows us to perform several self-consistency check on the signals provided by redundant detection, which matters for the diagnosis of cycle-slips and the discrimination of their origin. Note that the results reported in [13–15] were proof-of-principle experiments with both fiber ends located at the same laboratory (the local site and the remote one are at the same place), and the two lasers to be compared were actually the same laser.

In this report, we present our experimental work on a fiber link of 2×43 km connecting SYRTE (Système de Références Temps-Espace) to LPL (Laboratoire de Physique des Lasers) through a dedicated pair of fibers over an urban area. In the frame of the forthcoming optical clock comparison to be accomplished between the National Physical Laboratory (NPL) in the UK and SYRTE in France, we were seeking for a robust scheme with ultimate stability and accuracy [17, 18]. We investigated on that purpose a hybrid architecture, using active compensation on one of the fiber and using local two-way setup on the second one [19]. We are reporting here the first experimental result of optical frequency comparison by LTW on an installed fiber network and in a realistic implementation, to our knowledge. We investigate the impact of ultra-stable laser instability on the fiber link noise floor by using two independent laser sources for the LTW comparison. The limiting factors of the frequency comparison capability of LTW have been rigorously investigated, and their contribution to the relative frequency instabilities are presented and discussed. Finally, we discuss the accuracy of the frequency transfer and show that the hybrid setup allows optical clock comparison to $1 \times 10^{-20}$ level.

## 2. Schematic overview of experimental methods for optical frequency comparison

The experimental scheme we implement for optical frequency transfer and comparison is shown in Fig. 1. The experiments were performed using a pair of 43-km-long dedicated fiber from the urban network [20–22], which links SYRTE and LPL, two laboratories in the urban area of Paris. The two fibers are denoted by fiber-1 and fiber-2. On fiber-1 we built a regular setup using active noise compensation (ANC) of the fiber noise. On fiber-2 we built a local two-way setup as introduced in [15]. We used two independent ultra-stable lasers operated at 1.5 micron, with a sub-Hz linewidth, and a typical frequency drift of about 1 Hz/s. We used a fiber interferometric setup, splitting Laser1 light into two branches, and followed by optical couplers in order to implement Michelson-like interferometers. The interferometric setup at SYRTE was realized with spliced components and placed in an aluminum box, surrounded by thick thermal insulator foam. At the output of the coupler, we set up the fiber pigtailed acousto-optic modulator (AOM1) to feedback for the compensation of the noise. The light was then injected into the long-haul dedicated fiber connecting the two laboratories. The light at the remote end was frequency-shifted by the AOM2 (constant shift), and part of this light was reflected back by the Faraday mirror FM2. The fiber noise ($\phi_{NI}$) was measured after a round trip with a photodiode PD1 by the beat note between the round-trip signal and the light reflected by the Faraday mirror FM1 at SYRTE. The beat signal was tracked and processed to actively compensate the fiber noise by acting on the carrier frequency of the acousto-optic modulator AOM1. It can be shown that the local optical phase ($\phi_I$) at FM1 is copied to the remote Faraday mirror FM2 with this ANC set-up as described in Appendix. These two points are shown in Fig. 1 as blue filled circles. For the local interferometric setup at SYRTE, the sensitivity of the differential optical phase to temperature, as defined in [15], was measured to be 7 fs/K. The frequency transfer performance on this link was reported in [20–22].

In order to check the relative frequency stability and accuracy of the ANC setup, we cascaded the two 43-km fibers and formed an 86-km-long fiber loop as described in [20–22]. The frequency of the beat-note signal between the local laser and the round-trip one was measured simultaneously by two dead-time-free counters [23] (one in Π−type and the other in Λ−type [24]) with a gate time of 1 s. There was no cycle slip in the frequency data during the total measurement time of 66 691 s. The fractional frequency instabilities of ANC transfer method in terms of Allan deviation [24] are shown in Fig. 2. Uncompensated 86-km fiber noise is also shown with a light gray solid line. The overlapping Allan deviation (ADEV) and the modified Allan deviation (MDEV) calculated from the Π−type counter data start from $1.4 \times 10^{-15}$ at 1 s and decrease with slope of −1 and −3/2, respectively, which is expected for white phase noise. The ADEV and the MDEV calculated from the Λ−type counter data start from $2.4 \times 10^{-17}$ at 1 s and decrease to $1 \times 10^{-20}$ and to $6 \times 10^{-21}$ at 10 000 s, respectively. The MDEVs of the data recorded using Π−type and Λ−type converge to the same value with good agreement. This limit is attributed to the phase variation related to the temperature fluctuation of the residual fiber-length-mismatch in the interferometer [15]. The mean frequency offset from the expected frequency of the remote output was calculated with the total Λ−type counter data to be $8 \times 10^{-21}$ with a statistical uncertainty given by $1 \times 10^{-20}$ (the long-term overlapping ADEV at 10 000s with Λ−data [25–27]). It is consistent with the mean frequency offset of $1.2 \times 10^{-20}$ of the total Π−type counter data and its statistical error of $1.7 \times 10^{-20}$ (calculated as the relative standard deviation divided by the length of the consecutive segment in case of white phase noise [14, 28]). This shows the reliability of this uncertainty value. Accordingly, it can be concluded that the optical phase of the local laser has been transferred to the 86-km link output end with an accuracy $< 2 \times 10^{-20}$ with ANC method and that no frequency bias has been observed within the statistical uncertainty of the data set.

In the hybrid frequency comparison setup in Fig.1, part of the light reaching the remote light was extracted and re-injected into the second fiber connecting the two laboratories. The second part of the setup is depicted on the lower part of Fig.1. On each of the two sides there was an optical coupler realizing a strongly unbalanced Michelson interferometer and an AOM used to apply a constant frequency shift to the laser light. A polarization controller and Faraday mirrors were used so that the beat note signals on PD3 and PD4 were optimized. A bi-directional erbium-doped fiber amplifier (b-EDFA) was used at the remote site to compensate for the transmission loss of Laser1. As described in [15], two signals were generated in each photodiode; one for the frequency difference between the local laser and the laser transmitted from the remote site, and the other for the fiber noise measurement. The fiber noise $\phi_{N2}$ could be measured at the local PD4 by using the round-trip signal of the local laser, if $\phi_{N2}$ was assumed to be the same in both up-link and down-link transmission. Alternatively, $\phi_{N2}$ could also be measured at the remote PD3 by using the round-trip signal of the remote laser in the same way. Detailed description of the signal processing is given in next section.

The real local measurement (LM) of the frequency difference of the two lasers was made by a photodiode (PD5) at the local site to compare with the LTW measurement.

With CTW method, the fiber noise $\phi_{N2}$ is not measured but eliminated from the frequency difference signal by combining the two beat-note signals recorded at each end sites on both local PD4 and remote PD3. As the effect of $\phi_{N2}$ has opposite signs in the two signals from PD3 and PD4, it can be eliminated by post-processing the two measurement data sets. The local data and the remote data should be measured synchronously [13]. Compared with CTW method, LTW method uses only the local measurement data, thus there is no need for data exchange between the local and the remote sites for a post-processing nor for synchronization of the data acquisition timing. However, noise rejection can be up to 4-times lower in the power spectral density in LTW since it is limited by the propagation time, which is 2 times more for the round-trip signal than for the CTW [15]. In addition, LTW is more sensitive to the fiber attenuation, since it requires a round-trip propagation in the same fiber, whereas CTW only requires one-way propagation.

In order to compare independent laser sources at the remote site, we used one fiber to transfer the ultra-stable laser light from SYRTE to LPL with an ANC setup as shown in Fig. 1, and we used the LTW setup being injected with a second independent ultra-stable laser. In this way, we used the second fiber to perform the LTW comparison between two independent ultra-stable lasers. We performed also a second series of experiments with only one ultra-stable laser. In that case, the instabilities arising from the ultra-stable laser frequency fluctuations were largely rejected, and the interferometric noise could be studied with deeper insights. All frequencies were recorded with dead-time free frequency counters [23].

Note that this hybrid configuration not only enables us to study the LTW performance but it is very convenient to transfer an ultra-stable laser signal to the remote site with a real-time evaluation of the transfer accuracy and stability thanks to the LTW signal at remote site. Moreover, this performance can be further checked with the LTW at local site and CTW, thus showing that this hybrid fiber link design constitutes a very reliable technique for frequency transfer.

### 3. Local two-way experiment

LTW measurement at the local site was done by the two signals on PD4; $\phi_{PD4_A} = (\phi_{L1} + \phi_{N2}) - \phi_{L2}$ measures the fiber-noise-uncompensated phase difference between the two lasers, of respective phases $\phi_{L1}$ and $\phi_{L2}$, and $\phi_{PD4_B} = 2\phi_{N2}$ measures the round-trip fiber noise. We used heterodyne detection technique to generate all the signals on a single photodiode. We engineered the frequency map in such a way that the signals were far enough apart to be easily filtered and separated. Each filtered signal was amplified and tracked with a tracking oscillator with a typical bandwidth of 100 kHz. The compensated LTW phase difference $\phi_{LTW}^{local}$ measured at the local site is given by [15]

$$\phi_{LTW}^{local} = \phi_{PD4_A} - \frac{\phi_{PD4_B}}{2} = \phi_{L1} - \phi_{L2} . \tag{1}$$

It is worthwhile to note that $\phi_{LTW}^{local}$ is measured only with the local measurements by PD4.

Likewise, LTW measurement is also possible at the remote site by using the two signals on PD3; $\phi_{PD3_A} = \phi_{L1} - (\phi_{L2} + \phi_{N2})$ and $\phi_{PD3_B} = 2\phi_{N2}$. The compensated LTW phase difference measured at the remote site $\phi_{LTW}^{remote}$ is given by

$$\phi_{LTW}^{remote} = \phi_{PD3_A} + \frac{\phi_{PD3_B}}{2} = \phi_{L1} - \phi_{L2} . \tag{2}$$

Meanwhile, CTW measurement is also possible with this setup. The compensated CTW phase difference $\phi_{CTW}$ is given by

$$\phi_{CTW} = \frac{\phi_{PD4_A} + \phi_{PD3_A}}{2} = \phi_{L1} - \phi_{L2} . \tag{3}$$

It is noted in Eq. (3) that $\phi_{CTW}$ measurement requires a synchronized data set measured at both the local and the remote sites [13].

To test the performance of the LTW method, the $\phi_{LTW}^{local}$ measurement was compared with the real local measurement $\phi_{LM}$ by using the frequency data simultaneously obtained with a $\Pi$−type counter and a $\Lambda$−type counter with a gate time of 1 s. One has $\phi_{LM} = \phi'_1 - \phi'_2$ where $\phi'_1$ ($\phi'_2$) is the phase of Laser1 (Laser2) at PD5. The phase difference $\phi_{LTW}^{local} - \phi_{LM}$ should be zero in the limit of a perfect fiber noise rejection and a perfect experimental setup. The fiber noise rejection is limited by the propagation delay, inducing a small difference between the forward and backward propagation noise [6, 15]. Other effects, such as Sagnac effect or polarization effects can also limit the fiber noise rejection, which is effective only for the fluctuations that are equal for the forward and backward propagation. The measured phase evolution of $\phi_{LTW}^{local} - \phi_{LM}$ in 100 000 s is shown in Fig. 3(a) as a thick black line. This phase evolution is attributed to three major contributions, which will be described below.

It can be easily shown that, with this ANC set-up, any linear laser drift is corrected by the action of the phase-locked loop (PLL) and thus the transferred remote phase is copying the local phase $\phi_1$ as if there was no time delay due to the propagation (see Appendix). However, in the case of LTW, the drift of the frequency difference between the two lasers affects the phase of $\phi_{LTW}^{local} - \phi_{LM}$ because of the time delay $\tau$ of the light propagation between the local and the remote site. According to the derivation in Appendix, the first component of the $\phi_{LTW}^{local} - \phi_{LM}$ phase evolution due to the drift of the frequency difference is given by

$$\phi_{drift} = -2\pi\tau\left[\left(v_1(t) - v_2(t)\right) - \left(v_1(t_0) - v_2(t_0)\right)\right], \tag{4}$$

where $v_1(t)$ and $v_2(t)$ are the frequencies of Laser1 and Laser2, respectively, at the entrance of the local interferometer at time $t$. $t_0$ is the moment when the phase measurement was started.

The second component of the phase evolution of $\phi_{LTW}^{local} - \phi_{LM}$ is due to the length mismatch $\delta L_{local} = (L_{14} + L_{16} - L_{15}) + (L_{11} + L_{12} - L_{13})$ in the local interferometer and is given by

$$\phi_{local} = 2\pi v(\delta L_{local})\gamma\left[T_{local}(t) - T_{local}(t_0)\right], \tag{5}$$

where $\gamma$ is a phase-temperature coefficient of silica fiber, which has a value of 37 fs/(K·m) for an optical carrier at 194.4 THz and at 298 K [15], and $T_{local}(t)$ is the temperature of the local interferometer. This contribution cancels out if $L_{13} = L_{11} + L_{12}$ (leading to $\phi'_1 = \phi_1$) and $L_{15} = L_{14} + L_{16}$. Similarly, the last component of the phase evolution of $\phi_{LTW}^{local} - \phi_{LM}$ is due to the length mismatch $\delta L_{remote} = L_{22} - L_{21} - L_{23}$ in the remote interferometer and is given by

$$\phi_{remote} = 2\pi v(\delta L_{remote})\gamma\left[T_{remote}(t) - T_{remote}(t_0)\right] \tag{6}$$

where $T_{remote}(t)$ is the temperature of the remote interferometer. This contribution cancels out when $L_{22} = L_{21} + L_{23}$. In the case of a common laser phase $\phi_{L1} = \phi_{L2}$, it would result in $\phi_1 = \phi_2$ where $\phi_2$ is the phase of Laser2 on FM3 (see Fig. 1).

$\phi_{drift}$ is shown with a thin orange line in Fig. 3(a) by use of Eq. (4) and the frequency data from the real local measurement on PD5. The temperature data were measured simultaneously with 5 s interval both at local interferometer box and at the remote one. $\phi_{local}$ and $\phi_{remote}$ are shown in Fig. 3(a) with a thin green line and a thin blue line, respectively. It is expected that there is a time delay between the measured temperature and the real temperature of the fiber interferometer due to the heat transfer time. These time delays were obtained by the correlation analysis between the phase evolution and the temperature measurement, yielding 2300 s for the local interferometer and 105 s for the remote interferometer. The temperature measurement data were shifted by these delay times. Then $\delta L_{local}$ and $\delta L_{remote}$ were determined to be 0.15 m and 0.35 m, respectively, from the multi-linear regression of the phase evolution with the local and the remote temperature data as the independent variables. One can see in Fig. 3(a) that the phase evolution of $\phi_{LTW}^{local} - \phi_{LM}$ can be nicely explained by the sum of phase errors from these three major components (thick pink line).

The fractional frequency instabilities in terms of Allan deviations in Fig. 3(b) were obtained using these phase evolution data. In Fig. 3(c) magnified view of MDEV for $\phi_{LTW}^{local} - \phi_{LM}$ and the instability contributions from the three major phase error sources are also shown; $\phi_{drift}$ with an orange line, $\phi_{local}$ with a green line, and $\phi_{remote}$ with a blue line. ADEV and MDEV calculated from the Π−type counter data start from $5 \times 10^{-16}$ at 1 s and decrease with slope of −1 and −3/2, respectively, which is expected for white phase noise. MDEV calculated from the Λ−type counter data starts from $1.4 \times 10^{-16}$ at 1 s and is limited by the contribution by $\phi_{remote}$ from 400 s to 2000 s (due to the room temperature control at the remote site), and then by that of $\phi_{drift}$ after 10 000 s, resulting in the minimum instability to be $8 \times 10^{-20}$ at 4000 s. The expected frequency instability of LTW measurement, when the three major components of the phase error are suppressed, can be calculated by the residual phase given by

$$\phi_{residual} = \phi_{LTW}^{local} - \phi_{LM} - \phi_{drift} - \phi_{local} - \phi_{remote} \tag{7}$$

This result is shown with a purple dashed line and the ultimate instability would be $2 \times 10^{-20}$ at 20 000 s with an experimental optimization. The ultimate accuracy of the frequency comparison with experimental optimization was estimated by calculating the mean value of the Λ−type frequency data of $\phi_{residual}$ to be $6 \times 10^{-21}$ with a statistical uncertainty of $3 \times 10^{-20}$ (the long-term overlapping ADEV at 20 000s with Λ−data [25−27]).

Next, we investigated the LTW optical frequency comparison by using the same laser at both sites. This configuration gives an experimental simulation when the drift of each laser is actively removed. $\phi_{drift}$ is zero in this case, and the phase evolution is expected to be given by the remaining two phase error sources. The measured phase evolution of $\phi_{LTW}^{local} - \phi_{LM}$ (thick black line), the estimated phase error due to $\phi_{local}$ (green line), the estimated phase error due to $\phi_{remote}$ (blue line), and the sum of the contributions from two major phase error sources (thick pink line) are shown in Fig. 3(d). The overall phase evolution of $\phi_{LTW}^{local} - \phi_{LM}$ is explained by the sum of major two phase error sources. $\phi_{residual}$ is shown in the lower part of Fig. 3(d) and is well within 1 rad. The fractional frequency instabilities of $\phi_{LTW}^{local} - \phi_{LM}$ and the two major phase error sources are shown in Fig. 3(e) and Fig. 3(f). ADEV and MDEV calculated from the Π−type counter data start from $4 \times 10^{-16}$ at 1 s and decrease with slope of −1 and −3/2, respectively, which is expected for white phase noise. MDEV calculated from the Λ−type counter data starts from $5 \times 10^{-17}$ at 1 s and is limited by the contribution by $\phi_{remote}$ around 400 s (due to the room temperature control at the remote site), resulting in the minimum instability to be $2.4 \times 10^{-20}$ at 10 000 s. The expected instability of $\phi_{residual}$ (purple dashed line) indicates that ultimate instability would be $1.2 \times 10^{-20}$ at 10 000 s if the remaining two major error sources were suppressed with fiber length optimization. The ultimate accuracy of the frequency comparison with experimental optimization was estimated by calculating the mean value of the Λ−type frequency data of $\phi_{residual}$ to be $-4 \times 10^{-21}$ with a statistical uncertainty of $3 \times 10^{-20}$ (the long-term overlapping ADEV at 10 000s with Λ−data [25−27]).

## 4. LTW with no remote length-mismatch using a partial Faraday mirror

In this section, we implement a scheme which automatically suppresses the length mismatch at the remote site. The experimental setup for this section is shown in Fig. 4. We insert a partial Faraday mirror (p-FM) at the remote setup, which enables a perfect match of the fiber length in the remote interferometer. When the setup in Fig. 4 is compared with that in Fig. 1, it can be seen that the action of the p-FM is equivalent to overlapping the two remote Faraday mirrors in Fig. 1. Thus, $\delta L_{remote}$ is zero in this case, and the phase evolution is expected to be given by the remaining two phase error sources; $\phi_{drift}$ and $\phi_{local}$.

The measured phase evolution of $\phi_{LTW}^{local} - \phi_{LM}$ (thick black line), the estimated phase error due to $\phi_{drift}$ (orange line), the estimated phase error due to $\phi_{local}$ (green line), and the sum of the contributions from the two major phase error sources (thick pink line) are shown in Fig. 5(a). The phase evolution of $\phi_{LTW}^{local} - \phi_{LM}$ is nicely explained by the sum of the two major phase error sources, and it is free from the remote temperature fluctuation.

The fractional frequency instabilities of $\phi_{LTW}^{local} - \phi_{LM}$ and the two major phase error sources are shown in Fig. 5(b) and Fig. 5(c). ADEV and MDEV calculated from the $\Pi$−type counter data start from $5 \times 10^{-16}$ at 1 s and decrease with slope of $-1$ and $-3/2$, respectively, which is expected for white phase noise. MDEV calculated from the $\Lambda$−type counter data starts from $7 \times 10^{-17}$ at 1 s and is limited by the contribution by $\phi_{drift}$ after 2000 s, resulting in the minimum instability to be $6 \times 10^{-20}$ at 2000 s. The expected instability of $\phi_{residual}$ indicates that ultimate instability would be $9 \times 10^{-21}$ at 5000 s if the phase errors due to $\phi_{drift}$ and $\phi_{local}$ are suppressed. The ultimate accuracy of the frequency comparison with experimental optimization was estimated by calculating the mean value of the $\Lambda$−type frequency data of $\phi_{residual}$ to be $6 \times 10^{-21}$ with a statistical uncertainty of $2 \times 10^{-20}$ (the long-term overlapping ADEV at 10 000s with $\Lambda$−data [25−27]).

Next, the same laser was used at both sites with no fiber-length mismatch at the remote interferometer in order to investigate the performance limit of the LTW method. Thus, both $\phi_{drift}$ and $\phi_{remote}$ were zero in this case.

The measured phase evolution of $\phi_{LTW}^{local} - \phi_{LM}$ (thick black line) and the estimated phase error due to $\phi_{local}$ (green line) are shown in Fig. 5(d). The result was free from the remote temperature fluctuation when we see the temperature data in Fig. 5(e). The phase evolution due to $\phi_{local}$ was less than 0.2 rad because the temperature variation of the local interferometer was relatively small. The phase evolution of $\phi_{LTW}^{local} - \phi_{LM}$ is not explained by $\phi_{local}$, as shown below in Fig. 5(f), $\phi_{local}$ being one order of magnitude below. It indicates that there is another small unknown residual phase error sources at this instability level of $1 \times 10^{-20}$. It corresponds to a phase fluctuation signature also present in the error signal (Fig. 5(d)). It could be attributed to a limited rejection of the laser drift or of the fiber noise, eventually due to a non-reciprocal fiber noise, as for instance power or polarization effects, which is not cancelled with the ANC or the two-way set-up, or instabilities of the electronic detection setup. Further investigations are needed to elucidate the physical effects at work at this very low level of instability.

The fractional frequency instabilities of $\phi_{LTW}^{local} - \phi_{LM}$ and the estimated phase error due to $\phi_{local}$ are shown in Fig. 5(f). ADEV and MDEV calculated from the Π−type counter data start from $6 \times 10^{-16}$ at 1 s and decrease with slope of −1 and −3/2, respectively, which is expected for white phase noise. MDEV calculated from the Λ−type counter data starts from $1 \times 10^{-16}$ at 1 s, resulting in the minimum instability to be $3 \times 10^{-20}$ at 4000 s. Although the local temperature measurement happened to be stopped at 8 200 s, one can see that there was no bump due to the room temperature effects on the length mismatch at the remote site in Fig. 5(f), as compared to Fig. 3 (e) and (f). The tiny bump around 500 s in Fig. 5(f) results from the residual phase fluctuations discussed above. It is expected that instability level of $1 \times 10^{-20}$ would be possible with a longer integration time. The accuracy of the frequency comparison was estimated by calculating the mean value of the Λ−type frequency data of $\phi_{LTW}^{local} - \phi_{LM}$ to be $2 \times 10^{-20}$ with a statistical uncertainty of $6 \times 10^{-20}$ (the long-term overlapping

ADEV at 4 000s with Λ−data [25−27]). It should be noted that this relatively large uncertainty was due to the short averaging time.

Before we conclude, we introduce one more useful information that can be derived using the hybrid fiber link demonstrated in this article: the expected performance of a uni-directional two-way [14], where two separate fiber are used for uplink and downlink, can be estimated, which is shown in Fig. 6. This was obtained by comparing the fiber noise in fiber-1 with ANC setup and that in fiber-2 with LTW setup. One can see that the frequency comparison at the level of $7 \times 10^{-18}$ at 10 000 s would be possible with uni-directional two-way measurement. The accuracy of the frequency comparison was estimated by using the Λ−type frequency data, and found to be to be $1.0 \times 10^{-17}$ with a statistical uncertainty of $1.8 \times 10^{-17}$. This is a confirmation of the experimental data reported in [14]. Strikingly, it shows that atomic fountains can be compared in real time with such a link configuration at mid-range. This can give a convenient option when the bi-directional propagation is not available, or when a trade-off between performance and cost has to be made.

## 5. Conclusions

We showed the experimental result of a local two-way optical frequency comparison over a 43-km-long urban fiber network using two independent lasers at each of the local and the remote site. The local two-way method does not need any synchronization between the remote and local instruments because all the required data are obtained locally. Even real-time processing can be possible by using RF devices such as tracking oscillators and frequency mixers for the operations of summation and division which are required to obtain the local two-way result in Eq. (1). Three limiting factors of the LTW comparison (the drift of the frequency difference between the two lasers, the length mismatches in the local interferometer and in the remote interferometer) have been investigated, showing that the fractional frequency instability level of $1 \times 10^{-20}$ at about 10 000 s is possible with the LTW method. It is noted that this is the first time that the laser instability contribution is evidenced experimentally. We also introduced a simple scheme of LTW with no length mismatch at the remote site by using a partial Faraday mirror. We showed in this article that hybrid scheme can provide very useful information on fundamental limits of optical fiber links, and therefore enhance the transfer capabilities and the self-diagnosis abilities.

## Appendix

In this section, we derive Eq. (4) that describes the phase error in LTW experiment due to the drift of the frequency difference between the two lasers. Let us assume that $\nu_{rem}(t)$ and $\nu_2(t)$ are the frequencies of the remote laser and the local laser, respectively, at the entrance of respective interferometer at time $t$, and the frequencies of AOM3 and AOM4 in Fig. 1 are $f_3$ and $f_4$, respectively. $\tau$ is the time delay of the light propagation between the local and the remote site. Three optical signals which enter PD4 are from the remote laser, the local laser, and the round-tripped local laser, whose frequencies are given by $\nu_{rem}(t-\tau) + f_3 + f_4$, $\nu_2(t)$, and $\nu_2(t-2\tau) + 2f_3 + 2f_4$.

We will first show that the remote frequency is equal to the laser frequency at the same time, with no delay, due to the ANC. For the sake of simplicity, we suppose that the fiber noise is negligible. In that case, the phase at remote end of fiber-1 is $\phi_{remote}(t) = \phi_{L1}(t-\tau) + \phi_C(t-\tau)$ where $\phi_C(t-\tau)$ is the correction applied at link input end. For ergodic and stationary processes this correction is given by $\phi_{L1}(t) - (\phi_{L1}(t-2\tau) + \phi_C(t-2\tau) + \phi_C(t)) = 0$. In first order approximation, we have $\phi_C(t-2\tau) + \phi_C(t) = 2\phi_C(t-\tau)$. It results in $\phi_{remote}(t) = \phi_{L1}(t-\tau) - (1/2)(\phi_{L1}(t-2\tau) + \phi_{L1}(t)) = \phi_{L1}(t)$ at first order. Thus, we have $\nu_{rem}(t-\tau) = \nu_1(t-\tau)$: at first order approximation, the laser drift is corrected by the PLL.

The two signals of LTW on PD4 are the fiber-noise-uncompensated phase difference between the two lasers and the round-trip fiber noise. The frequency of the first signal is given by $\nu_{PD4A} = -\nu_2(t) + [\nu_1(t-\tau) + f_3 + f_4]$ and that of the second signal is given by $\nu_{PD4B} = -\nu_2(t) + [\nu_2(t-2\tau) + 2f_3 + 2f_4]$.

An approximate expression for the frequency of the LTW signal can be given by use of the Taylor expansion $\nu(t - a\tau) = \nu(t) - a\tau \frac{d\nu(t)}{dt} + \frac{(a\tau)^2}{2}\frac{d^2\nu(t)}{dt^2} + O(\tau^3)$.

$$v_{PD\,4A} - v_{PD\,4B}/2 = -\frac{1}{2}v_2(t) + v_1(t-\tau) - \frac{1}{2}v_2(t-2\tau)$$

$$\approx -[v_2(t) - v_1(t)] + \tau\left[\frac{dv_2(t)}{dt} - \frac{dv_1(t)}{dt}\right] - \tau^2\left[\frac{d^2v_2(t)}{dt^2} - \frac{1}{2}\frac{d^2v_1(t)}{dt^2}\right] \quad (8)$$

Since $\frac{dv(t)}{dt}$ and $\frac{d^2v(t)}{dt^2}$ are order of 1 Hz/s and 1 Hz/s$^2$, and $\tau$ is about $2 \times 10^{-4}$ s with 43 km fiber link, the third term in Eq. (8) can be neglected. The first term is cancelled when $\phi_{LTW}^{local}$ is compared with $\phi_{LM}$. Thus,

$$\phi_{drift} = 2\pi\int \tau\left[\frac{dv_2(t)}{dt} - \frac{dv_1(t)}{dt}\right]dt = 2\pi\tau[(v_2(t) - v_1(t)) - (v_2(t_0) - v_1(t_0))].$$

## Funding

This work was partly funded from the EC's Seventh Framework Programme (FP7 2007–2013) under Grant Agreement No. 605243 (GN3plus), Action spécifique GRAM, and from the European Metrological Research Programme EMRP under SIB-02 NEAT-FT. The EMRP is jointly funded by the EMRP participating countries within EURAMET and the European Union. W.-K. Lee was supported partly by the Korea Research Institute of Standards and Science under the project "Research on Time and Space Measurements," Grant No. 16011007, and also partly by the R&D Convergence Program of NST (National Research Council of Science and Technology) of Republic of Korea (Grant No. CAP-15-08-KRISS).

**References:**

1. C. Lisdat, G. Grosche, N. Quintin, C. Shi, S.M.F. Raupach, C. Grebing, D. Nicolodi, F. Stefani, A. Al-Masoudi, S. Dörscher, S. Häfner, J.-L. Robyr, N. Chiodo, S. Bilicki, E. Bookjans, A. Koczwara, S. Koke, A. Kuhl, F. Wiotte, F. Meynadier, E. Camisard, M. Abgrall, M. Lours, T. Legero, H. Schnatz, U. Sterr, H. Denker, C. Chardonnet, Y. Le Coq, G. Santarelli, A. Amy-Klein, R. Le Targat, J. Lodewyck, O. Lopez, and P.-E. Pottie, "A clock network for geodesy and fundamental science," Nat. Commun. **7**, 12443 (2016).
2. K. Predehl, G. Grosche, S. M. F. Raupach, S. Droste, O. Terra, J. Alnis, Th. Legero, T. W. Hänsch, Th. Udem, R. Holzwarth, and H. Schnatz, "A 920-kilometer optical fiber link for frequency metrology at the 19th decimal place," Science **336**(6080), 441–444 (2012).
3. O. Lopez, A. Haboucha, B. Chanteau, Ch. Chardonnet, A. Amy-Klein, and G. Santarelli, "Ultra-stable long distance optical frequency distribution using the Internet fiber network," Opt. Express **20**(21), 23518–23526 (2012).
4. D. Calonico, E. K. Bertacco, C. E. Calosso, C. Clivati, G. A. Costanzo, M. Frittelli, A. Godone, A. Mura, N. Poli, D. V. Sutyrin, G. Tino, M. E. Zucco, and F. Levi, "High-accuracy coherent optical frequency transfer over a doubled 642-km fiber link," App. Phys. B, **117**(3), 979–986 (2014).
5. W. Yang, D. Li, S. Zhang, and J. Zhao, "Hunting for dark matter with ultrastable fibre as frequency delay system," Scientific Reports, **5**, 11469 (2015).
6. P. A. Williams, W. C. Swann, and N. R. Newbury, "High-stability transfer of an optical frequency over long fiber-optic links," J. Opt. Soc. Am. B **25**(8), 1284–1293 (2008).
7. A. Matveev, C. G. Parthey, K. Predehl, J. Alnis, A. Beyer, R. Holzwarth, Th. Udem, T. Wilken, N. Kolachevsky, M. Abgrall, D. Rovera, Ch. Salomon, P. Laurent, G. Grosche, O. Terra, Th. Legero, H. Schnatz, S. Weyers, B. Altschul, and T. W. Hänsch, "Precision measurement of the hydrogen 1S-2S frequency via a 920-km fiber link," Phys. Rev. Lett., **110**(23), 230801 (2013).
8. B. Argence, B. Chanteau, O. Lopez, D. Nicolodi, M. Abgrall, Ch. Chardonnet, C. Daussy, B. Darquié, Y. Le Coq, and A. Amy-Klein, "Quantum cascade laser frequency stabilization at the sub-Hz level," Nature Photon. **9**(7), 456–461 (2014).
9. C. Clivati, G. Cappellini, L. Livi, F. Poggiali, M. Siciliani de Cumis, M. Mancini, G. Pagano, M. Frittelli, A. Mura, G. A. Costanzo, F. Levi, D. Calonico, L. Fallani, J. Catani, and M. Inguscio, "Measuring absolute frequencies beyond the GPS limit via long-haul optical frequency dissemination," Opt. Express **24**(11), 11865–11875 (2016).
10. C. Clivati, G. A. Costanzo, M. Frittelli, F. Levi, A. Mura, M. Zucco, R. Ambrosini, C. Bortolotti, F. Perini, M. Roma, and D. Calonico, "A coherent fiber link for very long baseline interferometry," IEEE T. Ultrason. Ferr., **62**(11), 1907–1912 (2015).
11. S. M. Foreman, A. D. Ludlow, M. H. G. de Miranda, J. E. Stalnaker, S. A. Diddams, and J. Ye, "Coherent optical phase transfer over a 32-km fiber with 1 s instability at $10^{-17}$," Phys. Rev. Lett., **99**(15), 153601 (2007).
12. C. Ma, L. Wu, Y. Jiang, H. Yu, Z. Bi, and L. Ma, "Coherence transfer of subhertz-linewidth laser light via an 82-km fiber link," Appl. Phys. Lett. **107**(26), 261109 (2015).
13. C. E. Calosso, E. Bertacco, D. Calonico, C. Clivati, G. A. Costanzo, M. Frittelli, F. Levi, A. Mura, and A. Godone, "Frequency transfer via a two-way optical phase comparison on a multiplexed fiber network," Opt. Lett. **39**(5), 1177–1180 (2014).
14. A. Bercy, F. Stefani, O. Lopez, Ch. Chardonnet, P.-E. Pottie, and A. Amy-Klein, "Two-way optical frequency comparisons at $5 \times 10^{-21}$ relative stability over 100-km telecommunication network fibers," Phys. Rev. A **90**(6), 061802(R) (2014).
15. F. Stefani, O. Lopez, A. Bercy, W.-K. Lee, Ch. Chardonnet, G. Santarelli, P.-E. Pottie, and A. Amy-Klein, "Tackling the limits of optical fiber links," J. Opt. Soc. Am. B **32**(5), 787–797 (2015).
16. D. W. Hanson, "Fundamentals of two-way time transfer by satellite," *in Proceedings of 43rd Annual Frequency Control Symposium*, pp. 174-178 (1989).
17. ICOF (International Clock Comparisons via Optical Fiber), http://geant3plus.archive.geant.net/opencall/Optical/Pages/ICOF.aspx.

18. J. Kronjäger, G. Marra, O. Lopez, N. Quintin, A. Amy-Klein, W.-K. Lee, P.-E. Pottie, and H. Schnatz, "Towards an international optical clock comparison between NPL and SYRTE using an optical fiber network," presented at the 30th European Frequency and Time Forum, York, United Kingdom in 4–7 April 2016.
19. W.-K. Lee, F. Stefani, A. Bercy, O. Lopez, A. Amy-Klein, P.-E. Pottie, "Strengthening capability of optical fiber link with hybrid solutions", 8th Symposium on Frequency Standards and Metrology 2015, Poster E05, pp. 155, Potsdam, Germany, in 12–16 Oct. 2015.
20. H. Jiang, F. Kéfélian, S. Crane, O. Lopez, M. Lours, J. Millo, D. Holleville, P. Lemonde, Ch. Chardonnet, A. Amy-Klein, and G. Santarelli, "Long-distance frequency transfer over an urban fiber link using optical phase stabilization," J. Opt. Soc. Am. B **25**(12), 2029–2035 (2008).
21. F. Kéfélian, O. Lopez, H. Jiang, Ch. Chardonnet, A. Amy-Klein, and G. Santarelli, "High-resolution optical frequency dissemination on a telecommunications network with data traffic," Opt. Lett. **34**(10), 1573–1575 (2009).
22. H. Jiang, "Development of ultra-stable laser sources and long-distance optical link via telecommunication networks," Ph. D. thesis (2010), https://tel.archives-ouvertes.fr/tel-00537971.
23. G. Kramer and W. Klische, "Multi-channel synchronous digital phase recorder," In Proceedings of the 2001 IEEE International Frequency Control Symposium and PDA Exhibition, pp. 144–151 (2001).
24. S. Dawkins, J. McFerran, and A. Luiten, "Considerations on the measurement of the stability of oscillators with frequency counters," IEEE Trans. Ultrason. Ferroelectr. Freq. Control **54**(5), 918–925 (2007).
25. N. Chiodo, N. Quintin, F. Stefani, F. Wiotte, E. Camisard, C. Chardonnet, G. Santarelli, A. Amy-Klein, P.-E. Pottie, and O. Lopez, "Cascaded optical fiber link using the internet network for remote clocks comparison," Opt. Express **23**(26), 33927–33937 (2015).
26. A. Bercy, O. Lopez, P.‑E. Pottie, and A. Amy‑Klein, "Ultrastable optical frequency dissemination on a multi‑access fibre network," Appl. Phys. B **122**, 189 (2016).
27. S. M. F. Raupach, A. Koczwara, and G. Grosche, "Brillouin amplification supports $1 \times 10^{-20}$ uncertainty in optical frequency transfer over 1400 km of underground fiber," Phys. Rev. A **92**(2), 021801(R) (2015).
28. W.-K. Lee, D.-H. Yu, C. Y. Park, and J. Mun, "The uncertainty associated with the weighted mean frequency of a phase-stabilized signal with white phase noise," Metrologia **47**(1), 24 (2010).

**Figure Captions:**

Fig. 1. Experimental scheme for the simultaneous optical frequency transfer and comparison. PD, photodiode; OC, optical coupler; AOM, acousto-optic modulator; PLL, phase-locked loop; b-EDFA, bi-directional erbium-doped fiber amplifier; PC, polarization controller.

Fig. 2. Fractional frequency instabilities of active noise compensation in terms of Allan deviation with Π−type counter and Λ−type counter in an end-to-end (86 km) scheme.

Fig. 3. (a) Phase evolution of the local two-way comparison and that of the three major phase error sources, (b) fractional frequency instability of $\phi_{LTW}^{local} - \phi_{LM}$ in terms of Allan deviation, (c) MDEV of the three major phase error sources. Independent lasers were used at both sites in (a) ~ (c). (d) ~ (f) correspond to each case of (a) ~ (c), respectively, when the same laser was used at both sites.

Fig. 4. Experimental setup with no fiber-length-mismatch at remote site. p-FM, partial Faraday mirror; PD, photodiode; OC, optical coupler; AOM, acousto-optic modulator; PLL, phase-locked loop; b-EDFA, bi-directional erbium-doped fiber amplifier; PC, polarization controller.

Fig. 5. (a) Phase evolution of the local two-way comparison and that of the two major phase error sources when there is no fiber-length-mismatch at remote site ($\phi_{remote} = 0$), (b) fractional frequency instability of $\phi_{LTW}^{local} - \phi_{LM}$ in terms of Allan deviation, (c) MDEV of the two major phase error sources. Independent lasers were used at both site in (a) ~ (c). (d) and (f) correspond to the case of (a) and (b), respectively, when the same laser was used at both sites and $\phi_{remote} = 0$. (e) Temperature data of the local interferometer and the remote interferometer in case of (d) and (f).

Fig. 6. Estimation of the expected performance of a uni-directional two-way in terms of the fractional frequency instabilities of the difference of the fiber noise of each fiber with Π−type counter and Λ−type counter.

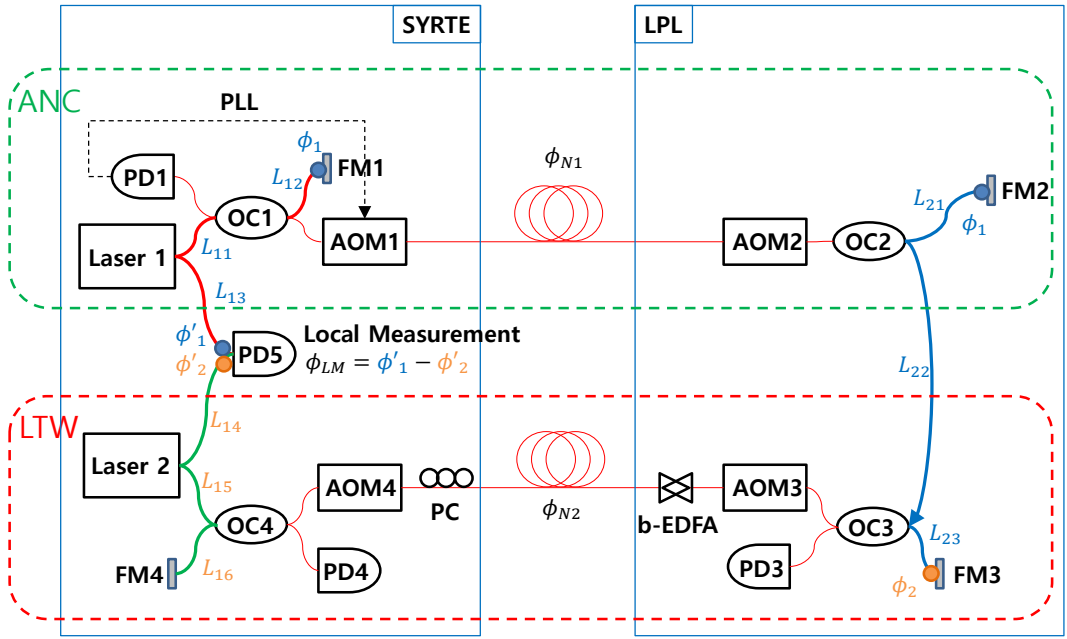

Fig. 1. Experimental scheme for the simultaneous optical frequency transfer and comparison. PD, photodiode; OC, optical coupler; AOM, acousto-optic modulator; PLL, phase-locked loop; b-EDFA, bi-directional erbium-doped fiber amplifier; PC, polarization controller.

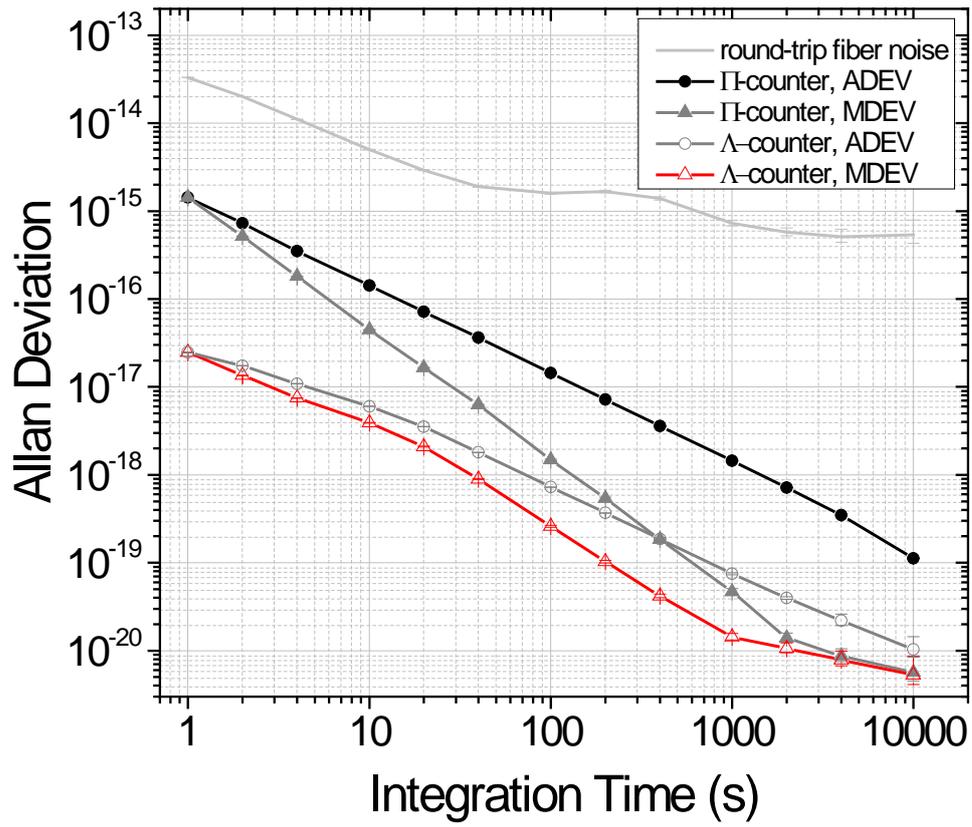

Fig. 2. Fractional frequency instabilities of active noise compensation in terms of Allan deviation with Π−type counter and Λ−type counter in an end-to-end (86 km) scheme.

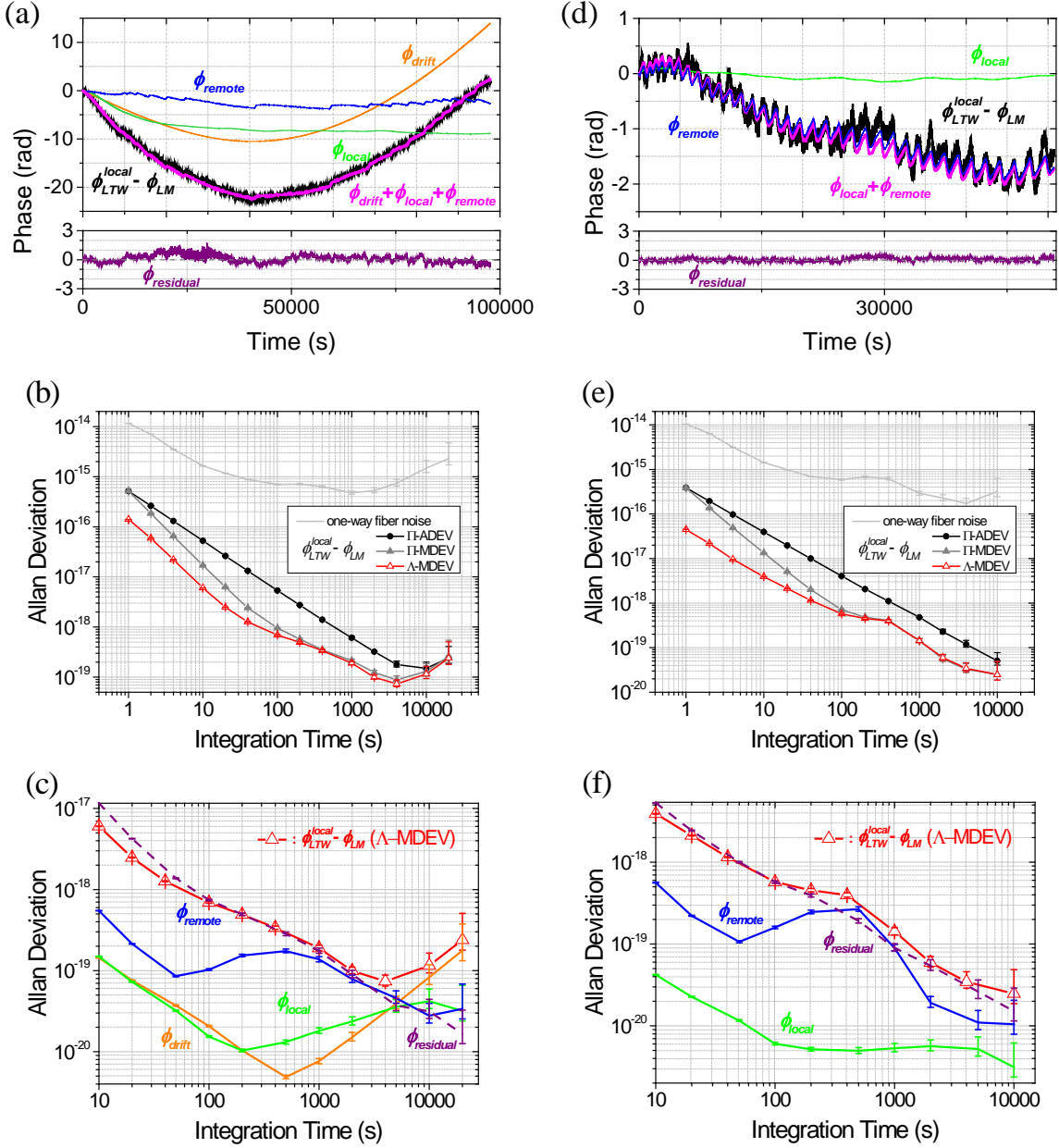

Fig. 3. (a) Phase evolution of the local two-way comparison and that of the three major phase error sources, (b) fractional frequency instability of $\phi_{LTW}^{local} - \phi_{LM}$ in terms of Allan deviation, (c) MDEV of the three major phase error sources. Independent lasers were used at both sites in (a) ~ (c). (d) ~ (f) correspond to each case of (a) ~ (c), respectively, when the same laser was used at both sites.

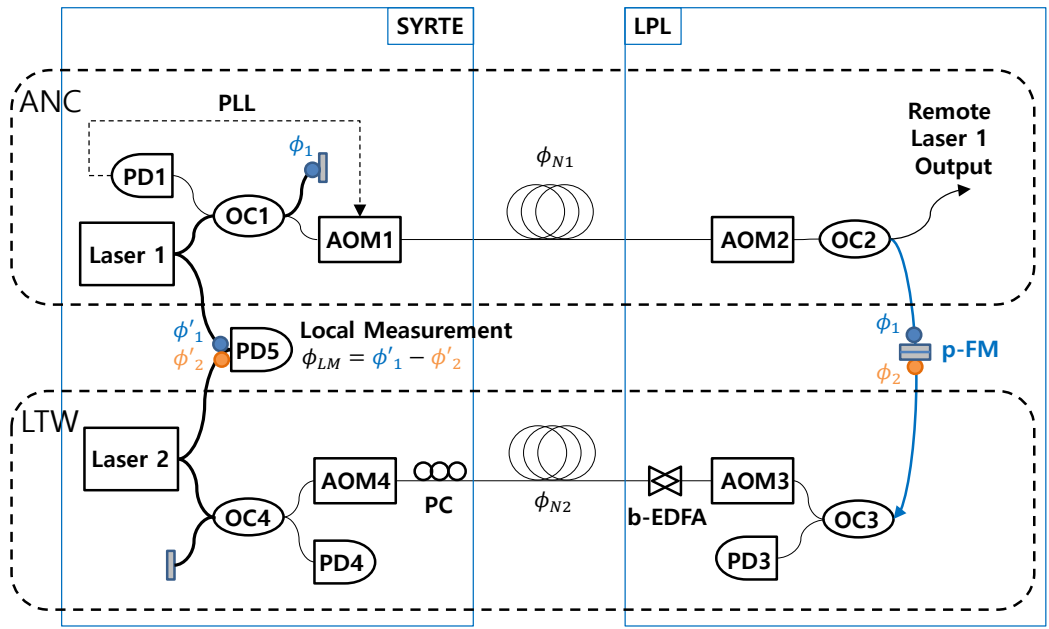

Fig. 4. Experimental setup with no fiber-length-mismatch at remote site. p-FM, partial Faraday mirror; PD, photodiode; OC, optical coupler; AOM, acousto-optic modulator; PLL, phase-locked loop; b-EDFA, bi-directional erbium-doped fiber amplifier; PC, polarization controller.

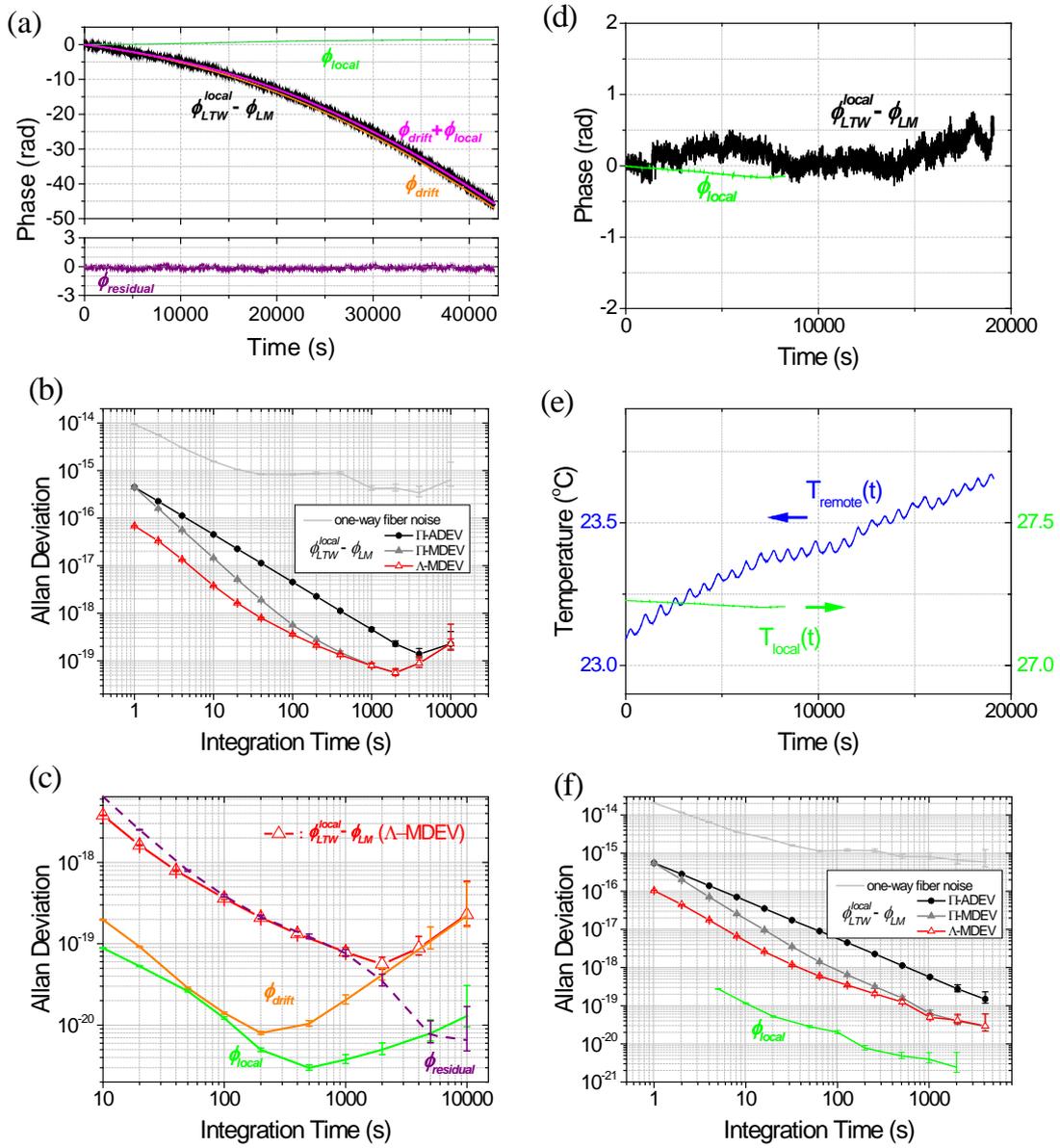

Fig. 5. (a) Phase evolution of the local two-way comparison and that of the two major phase error sources when there is no fiber-length-mismatch at remote site ($\phi_{remote} = 0$), (b) fractional frequency instability of $\phi_{LTW}^{local} - \phi_{LM}$ in terms of Allan deviation, (c) MDEV of the two major phase error sources. Independent lasers were used at both site in (a) ~ (c). (d) and (f) correspond to the case of (a) and (b), respectively, when the same laser was used at both sites and $\phi_{remote} = 0$. (e) Temperature data of the local interferometer and the remote interferometer in case of (d) and (f).

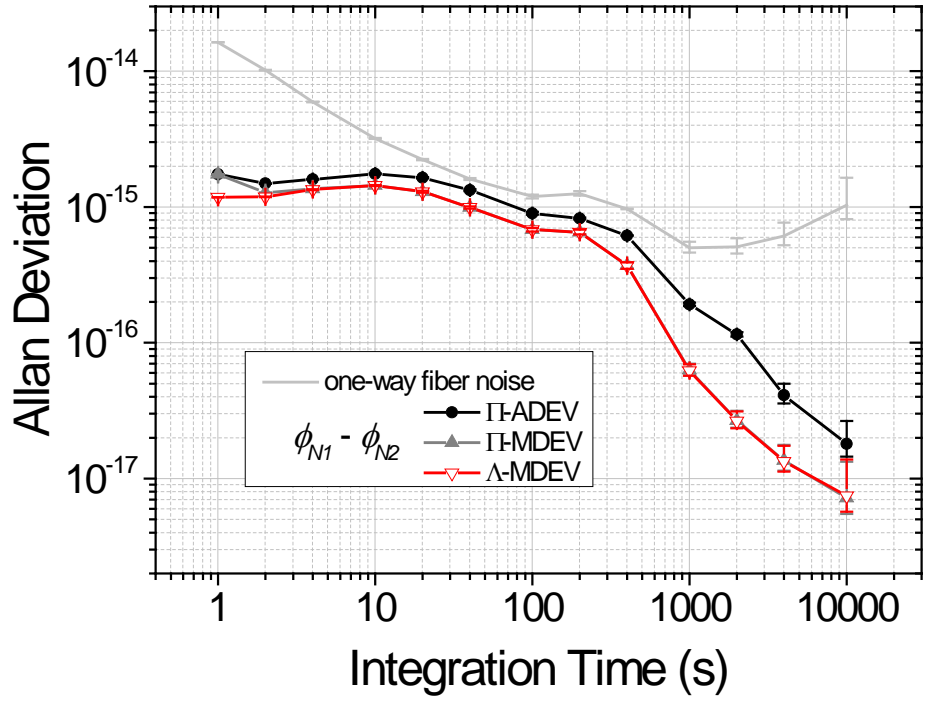

Fig. 6. Estimation of the expected performance of a uni-directional two-way in terms of the fractional frequency instabilities of the difference of the fiber noise of each fiber with Π−type counter and Λ−type counter.